\documentstyle[12pt]{article}
\input{psfig.sty}
\newlength{\dinwidth}
\newlength{\dinmargin}
\newlength{\extraspace}
\newlength{\extraspaces}
\def\simlt{\mathrel{\lower2.5pt\vbox{\lineskip=0pt\baselineskip=0pt
           \hbox{$<$}\hbox{$\sim$}}}}
\begin{document}
%
\setcounter{footnote}{1}
\renewcommand{\thefootnote}{\fnsymbol{footnote}}
\begin{flushright}
MADPH-98-1062\\
\end{flushright}
\vspace{14mm}
\begin{center}
\Large{{\bf 
Triple Gauge Boson Couplings \\
in \\
Rare $B$ and $K$ Decays}}
\end{center}
\vspace{5mm}
\begin{center}
{
Gustavo Burdman}\footnote{e-mail address: burdman@pheno.physics.wisc.edu}\\
*[3.5mm]
{\normalsize\it Department of Physics, University of Wisconsin,}\\ 
{\normalsize
\it Madison, WI 53706, USA}
\end{center}
\vspace{0.50cm}
\thispagestyle{empty}
\begin{abstract}
We examine 
the sensitivity of flavor changing neutral current (FCNC)
processes 
to anomalous triple gauge boson couplings. 
We show that in the non-linear 
realization of the electroweak symmetry breaking 
sector these processes are very sensitive to two 
CP conserving anomalous couplings.   
A clean separation of their  effects is 
possible in the next round of experiments probing $b\to s\gamma$ and 
$b\to s\ell^+\ell^-$ processes, as well as kaon decays such as 
$K^+\to\pi^+\nu\bar\nu$.    
The obtained sensitivity is found to be competitive with that 
of direct measurements at high energy colliders. In particular, for one 
of the $WWZ$ couplings the one-loop FCNC effects are enhanced by a 
logarithmic dependence on the scale of new physics.  
We also explore the potential signals of CP violating anomalous 
triple gauge boson couplings in rare $B$ decays.

\end{abstract}
\newpage
\setcounter{footnote}{1}
\setcounter{page}{1}

\section{Introduction} 
\vspace{-0.2cm}
The remarkable experimental success of the standard model (SM) 
suggests the possibility that at the weak scale  there may be no other 
dynamics or particle content. On the other hand, several questions 
remain unanswered within the SM framework and may require new dynamics 
in order to be addressed.  
Among these questions are the origin of electroweak symmetry breaking 
and of fermion masses. In principle, it could be argued that the energy 
scales
of the new dynamics related to these questions may be so large as to be 
irrelevant to  observables at the weak scale. However, it is known that 
the physics behind the Higgs sector and responsible for the breaking of the 
electroweak symmetry, cannot reside at scales much higher than 
a few TeV. Furthermore, it is possible that the origin of 
the top quark mass might be related to electroweak symmetry breaking. 
Thus, at least in some cases, the dynamics associated with new physics 
may not reside at arbitrarily high energies and there might be 
observable effects at lower energies. 
In cases where the underlying dynamics is not known or fully understood, 
the study of these non-decoupling effects is the realm of effective field
theory. 
The non-decoupling effects of the Higgs mechanism in the electroweak symmetry
breaking (EWSM) sector of the SM have been vastly studied in the 
literature~\cite{longhitano,appelquist,feruglio}. 
In order to write down the effective theory at energies well below the 
new physics scale $\Lambda$, all states with masses above $\Lambda$ must 
be integrated out. The result is an effective field theory for the gauge
bosons of the electroweak gauge group and the Nambu-Goldstone bosons (NGB)
associated with the spontaneous breaking of $SU(2)_L\times U(1)_Y$ 
down to $U(1)_{\rm EM}$. 
The effective theory at weak scale energies and below is in general 
non-renormalizable. However, it is possible to expand it in terms of 
the increasing dimension of the operators: the higher the dimension of the 
operator the higher the inverse power of $\Lambda$ suppressing its effects. 
Up to a given order (e.g. operators of dimension six, eight, etc. ) it is
possible to obtain a predictive effective theory.
The effects of the  physics above the scale $\Lambda$ are encoded
in the values of the coefficients of the higher-dimension operators.

We will concentrate on a scenario without scalars with masses below
$\Lambda$. This choice is motivated by the fact that the presence
of a light scalar is usually either accompanied by other new particles with 
masses of the order of the weak scale
(e.g. supersymmetry)  or allows for the scale of new physics to be 
very large~\cite{quiros}, thereby resulting in suppressed effects. 
This scenario is most appropriately described by a Higgs
sector with non-linear transformation properties~\cite{feruglio}. 
However, as we will see below, for the most part 
our results will be independent of this 
choice.
We will stress the relevance of using the non-linear realization when 
necessary. 

Important constraints on deviations from the SM through the coefficients of 
the 
effective Lagrangian of the EWSB sector come from electroweak observables at 
the weak scale. For instance, non-standard contributions to two-point 
functions
are severely constrained by oblique parameters as measured at the 
$Z$-pole~\cite{zpole}. 
Contributions to anomalous triple gauge-boson couplings (TGC) are bound 
by measurements of gauge boson production at LEPII~\cite{dkglep2}
and the Tevatron~\cite{dkgtev} as well
as by indirect measurements~\cite{sally,sally2}, whereas 
anomalous quartic couplings 
give one-loop contributions to oblique parameters. 
Finally, there is a set of operators in the EWSB sector that amounts to 
corrections to the NGB
propagators and that result in 
four-fermion operators coupling to fermion masses. These are not bound
at one-loop by oblique corrections, but by their contributions to vertices 
through top-quark loops~\cite{pich,bews}.
 
We are interested in evaluating the sensitivity of FCNC decay processes
at low energies, such as loop-induced $B$ and $K$ decays, to  
new dynamics in the EWSB sector residing above the scale $\Lambda$. 
These processes , such as $b\to s\gamma$, $b\to s\ell^+\ell^-$, 
$K\to\pi \nu\bar\nu$, etc., are affected in principle by all non-standard 
couplings of the gauge bosons  and the NGB. In practice, since 
oblique corrections are directly probed with high precision at the $Z$ pole, 
the corrections to two-point functions  are already highly constrained and 
will give no effect when included in the one-loop processes named above. 
Moreover, non-standard quartic gauge boson couplings do not enter in these
decays to leading order. Thus, we are left with two sources of deviations
from the SM expectations in these modes: corrections to the NGB
propagators and anomalous TGC. 
In Ref.~\cite{bews} the sensitivity of rare $B$ and $K$ decays to 
the corrections to NGB propagators was studied. It was there concluded that 
within the constraints imposed on the effective lagrangian parameters by 
the measurements of $R_b$, and $B$ and $K$ mixing large deviations from the 
SM were still possible in most FCNC decay modes, with the exception of 
$b\to s\gamma$ and $b\to d\gamma$. In this paper, we want to evaluate 
the sensitivity of these decay modes to anomalous TGC originated, through
the effective lagrangian, at energies above $\Lambda$. 

The effects of anomalous TGC in rare $B$ decays have been previously 
studied in the literature. 
For instance, the effects of the 
dimension four anomalous $WW\gamma$ coupling $\Delta\kappa_\gamma$ in $b\to s\gamma$ 
transitions
were first considered in~\cite{chia}, whereas this plus the dimension six 
coupling $\lambda_\gamma$ where studied in~\cite{numata,rizzo}. 
These plus the corresponding CP violating  couplings and their effects in the 
$b\to s\gamma$ branching fractions
were also considered in~\cite{he}. Finally, the anomalous $WWZ$ couplings 
and their effects in $b\to s\mu^+\mu^-$ were  studied in 
Ref.~\cite{baillie}. 
In this paper, we use the power counting of the non-linear realization 
of the EWSB sector to organize the anomalous 
TGC according to the dimension of the operator generating them in the 
effective theory.
This will identify the relevant anomalous TGC in scenarios
where the EWSB sector is strongly coupled. 
We will see that in these cases, FCNC transitions are very sensitive to one 
$WW\gamma$ and one $WWZ$ anomalous couplings, thus offering very well
defined constraints on the strongly coupled EWSB sector that are 
competitive, 
for these couplings, to those obtained at higher energies.  
We also add the constraints from present and future measurements of 
rare kaon decays such as 
$K^+\to\pi^+\nu\bar\nu$. 
Previous studies of the effects of anomalous TGC couplings 
in rare $K$ decays were done for $K\to\ell^+\ell^-$ decays~\cite{he2}, 
a mode largely 
affected by long distance contributions, and for $K^+\to\pi^+\nu\bar\nu$
by considering the effects of a parity violating anomalous TGC 
coupling~\cite{sally}. In this paper we study the effects of the two 
relevant couplings and put the effects in the context of the 
a specific scenario for EWSB and with the effects in $B$ decays. 
We complete the analysis 
by considering  the effects of CP violating TGC in rare $B$ decays,  
both in the rate as well as in CP asymmetries. 

The purpose of  this work is to evaluate the 
sensitivity of future $B$ and $K$ experiments
to anomalous TGC in the context of a strongly coupled EWSB sector. 
Although model-independent in nature, this context 
results in a hierarchy of anomalous TGC related to the 
power counting in the resulting 
effective theory. A complete treatment of the effects of this 
scenario in rare $B$ and $K$ 
decays is lacking in the literature. 
This forms part of a program started in 
Ref.~\cite{bews}, intended to explore the reach of sensitivity processes 
like the ones discussed
here to a strongly coupled EWSB sector.
It is possible that in a scenario like this one direct signals  
will not become available
until the CERN Large Hadron Collider (LHC) begins taking data.  
We also evaluate the competitiveness 
of these measurements with  
the direct measurements at higher energies to take place at the CERN-LEPII 
and the Fermilab-Tevatron colliders. 
We find these two approaches complementary largely due to the fact that 
the rare decay modes are selectively sensitive to a handful of 
anomalous TGC allowing  independent measurements of these couplings.  
In the next Section we review the non-linear realization of the 
effective lagrangian of the EWSB sector in relation to 
anomalous TGC. In Section~3 we compute the 
effects in rare $B$ and $K$ decays and 
we discuss the results and conclude in Section~4. 

\section{The Effective Lagrangian and Anomalous TGC}
In the absence of a light Higgs boson the symmetry breaking sector 
is represented by a non-renormalizable effective lagrangian
corresponding to the non-linear realization of the $\sigma$ model. 
The essential feature is the spontaneous breaking of the global symmetry
$SU(2)_L\times SU(2)_R\to SU(2)_V$. To leading order the interactions
involving the NGB associated with this mechanism and 
the gauge fields are described by
\begin{equation}
{\cal L}_{LO}=-\frac{1}{4}B_{\mu\nu}B^{\mu\nu}
-\frac{1}{2}{\rm Tr}\left[W_{\mu\nu}W^{\mu\nu}\right]
+\frac{v^2}{4}{\rm Tr}\left[D_\mu U^\dagger D^\mu U\right] ,
\label{lo}
\end{equation}
where $B_{\mu\nu}$ and  
$W_{\mu\nu}=\partial_\mu W_\nu-\partial_\nu W_\mu + 
ig\left[W_\mu,W_\nu\right]$
are the  the $U(1)_Y$ and $SU(2)_L$ field strengths respectively, 
the electroweak scale is $v\simeq 246$~GeV and 
the NGB enter through the matrices
$U(x)=e^{i\pi(x)^a\tau_a/v}$. The covariant derivative acting on $U(x)$
is given by $D_\mu U(x)=\partial_\mu U(x)+ igW_\mu (x) U(x) - 
\frac{i}{2}g'B_\mu (x) U(x)\tau_3$. To this order there are no free
parameters once the gauge bosons masses are fixed. The 
dependence on the dynamics underlying the strong symmetry breaking
sector appears at next-to-leading order. A complete set of 
operators at next to leading order includes one operator of 
dimension two and  operators of dimension four 
\cite{longhitano,appelquist}. 
The effective lagrangian to next to leading order 
is given by (see the Appendix for the expanded operator basis)
\begin{equation}
{\cal L}_{\rm eff.}={\cal L}_{LO} + 
{\cal L}'_1 +\sum_{i=1}^{19}\alpha_i{\cal L}_i ~~,
\label{lnlo}
\end{equation}
where ${\cal L}'_1$ is a
dimension two custodial symmetry violating term absent in the heavy
Higgs limit of the SM. If we restrict ourselves to CP invariant
structures, there remain fourteen operators of dimension four. 
As it was mentioned above, 
the coefficients of some of these operators are constrained by low
energy observables. For instance precision electroweak observables
bound the coefficient of ${\cal L}'_1$, which gives a contribution
to $\Delta\rho$. The combinations $(\alpha_1+\alpha_8)$ and $(\alpha_1
+\alpha_{13})$ contribute to the oblique parameters $S$ and $U$, defined
in~\cite{zpole}.
Corrections to the charged and neutral NGB propagators 
come from the operators ${\cal L}_{11}$ and ${\cal L}_{12}$ respectively.
Their effects in $B$ and $K$ FCNC processes were 
studied in Ref.~\cite{bews}.  
The coefficients $\alpha_2$, $\alpha_3$, $\alpha_9$ and $\alpha_{14}$
modify the TGC and are the object of our study. 

Imposing $CP$ conservation\footnote{We discuss possible effects 
from $CP$ violating anomalous TGC later in the paper.}, 
the most general form of the 
$WWN$ ($N=\gamma, Z$)
couplings can be written as~\cite{dieter}
\begin{eqnarray}
{\cal L}_{WWN}&=&g_{WWN}\Large\left\{i\kappa_N W_{\mu}^{\dagger}W_\nu 
N^{\mu\nu}
+ig_1^N \left(W_{\mu\nu}^{\dagger}W^\mu N^\nu -
W_{\mu\nu}W^{\dagger\mu} N^\nu\right) \right. \nonumber\\
& & \left. +g_5^V\epsilon^{\mu\nu\rho\sigma}(W^\dagger_\mu\partial_\rho 
W_\nu 
-W_\mu\partial_\rho W^\dagger_\nu)N_\sigma
+i\frac{\lambda_N}{M_W^2}W^\dagger_{\mu\nu}W^\nu_{~\lambda} 
N^{\nu\lambda}
\right\}~, 
\label{wwnc}
\end{eqnarray}
with the conventional choices being $g_{WW\gamma}=-e$ and 
$g_{WWZ}=-g\cos\theta$. 
In principle, there are six free parameters, since 
gauge invariance implies $\Delta g_1^\gamma=g_5^\gamma=0$. 
Making contact with the electroweak lagrangian (\ref{lnlo}), 
these parameters can be expressed in terms of the next-to-leading
order coefficients~\cite{holdom,appelquist}
\begin{eqnarray}
\Delta\kappa_\gamma\equiv\kappa_\gamma -1 &=& g^2(\alpha_2 
-\alpha_1+\alpha_3 -\alpha_8
+\alpha_9)  \nonumber\\
\Delta\kappa_Z\equiv\kappa_Z -1 &=& g^2(\alpha_3 -\alpha_8+ 
\alpha_9) +g'^2(\alpha_1-\alpha_2)  \nonumber\\
\Delta g_1^Z\equiv g_1^Z-1 &=& \frac{g^2}{\cos^2\theta_W}\alpha_3\nonumber\\
g_5^Z &=& \frac{g^2}{\cos^2\theta_W}\alpha_{14}\nonumber\\
\quad\quad\quad\lambda_\gamma=\lambda_Z&=&0~,
\label{contact}
\end{eqnarray}
where $g$ and $g'$ are the $SU(2)_L$ and $U(1)_Y$ gauge couplings 
respectively,
and the operator basis is the one defined in~\cite{longhitano}. 
As we see from the last line in (\ref{contact}), 
to this order in the energy expansion (\ref{lnlo}) we obtain
$\lambda_N=0$. These TGC get contributions from operators of dimension
six, suppressed by an extra factor of $(v^2/\Lambda^2)$.  
We are left with $\kappa_\gamma$, $\kappa_Z$, $g_1^Z$ and $g_5^Z$. 
Finally, when considering rare $B$ and $K$ decays, we can neglect the 
contributions from $\kappa_Z$ since they will be suppressed by powers of the 
squared of the external momenta over $m_Z^2$. 
Thus, in this approach, there are only three parameters
relevant at very low energies. The SM predictions for them 
are $\kappa_\gamma=g_1^Z=1$ and  $g_5^Z=0$.

\section{The Effects in FCNC Decays}

The presence of the 
anomalous TGC $\Delta\kappa_\gamma$, $\Delta g_1^Z$  and $g_5^Z$
will result in deviations from the SM in various FCNC $B$ and $K$ decays
\footnote{Charm FCNC decays are generally affected by large
long-distance contributions that tend to obscure the extraction 
of short distance physics. Although there are some exceptions to this 
statement,
the effects of anomalous TGC are  not among them~\cite{charm}.}. 
We first concentrate on rare $B$ decays, with focus on strategies to 
make use of large data samples for the various modes. 
We then present the constraints from $K^+\to\pi^+\nu\bar\nu$ measurements, 
and finally study the possible effects of  CP violating anomalous TGC.

\subsection{Rare $B$ Decays}

For the $b\to s\gamma$ and $b\to s\ell^+\ell^-$  transitions it is 
useful to cast the contributions of the anomalous couplings 
as shifts in the matching conditions at $M_W$ for the 
Wilson coefficient functions in the weak effective 
hamiltonian  
\begin{equation}
H_{\rm eff.}=-\frac{4G_F}{\sqrt{2}}\sum_{i=1}^{10} C_i(\mu)O_i(\mu),
\label{heff}
\end{equation}
with the operator basis defined in Ref.~\cite{heff}.
Of interest in our analysis are the electromagnetic penguin operator
\begin{equation}
{\cal O}_7 = \frac{e}{16\pi^2}m_b\; 
(\bar{s}_L\sigma_{\mu\nu}b_R)\;F^{\mu\nu}~,
\label{o7}
\end{equation}
and the four-fermion operators corresponding to the vector and axial-vector
couplings to leptons, 
\begin{equation}
{\cal O}_9 = \frac{e^2}{16\pi^2}\;(\bar{s}_L\gamma_\mu b_L)
(\bar{\ell}\gamma^\mu\ell)
\label{o9}
\end{equation}
and
\begin{equation}
{\cal O}_{10} = \frac{e^2}{16\pi^2}\;(\bar{s}_L\gamma_\mu b_L)
(\bar{\ell}\gamma^\mu\gamma_5\ell)~.
\label{o10}
\end{equation}

We first turn to the effects of $\Delta\kappa_\gamma$ in $b\to s\gamma$ and 
$b\to s\ell^+\ell^-$
transitions. This modification of the  $W^+ W^-\gamma$ coupling gives a 
shift in the one-loop  $b\to q\gamma$ vertex, with $(q=d,s)$. 
For $q=s$ this is given by
\begin{eqnarray}
\delta\Gamma_{\mu}^{b\to s\gamma}=i\frac{e}{4\pi^2}~\frac{G_F}{\sqrt{2}}
~V_{ts}^*V_{tb}
& &\hspace*{-0.5cm}\left\{\delta C_7(M_W)~m_b\bar{s}_L\sigma_{\mu\nu}b_Rq^\nu\right.\nonumber\\
& &\hspace*{-0.5cm}\left.+\delta C_9(M_W)~\bar{s}_L
\left(\not\!q q_\mu-q^2\gamma_\mu\right)b_L 
\right\}~, 
\label{bsgamma}
\end{eqnarray}
where $q_\mu$ is the photon four-momentum, 
only the top quark contributions are kept and  terms suppressed by 
$m_s/m_b$ have been neglected. 
These shifts in the Wilson coefficients at $M_W$ are given by 
\begin{eqnarray}
\delta C_7(M_W)&=& \frac{1}{2}\Delta\kappa_\gamma A_1(x_t)\label{dc7}\\
\delta C_9(M_W)&=& \Delta\kappa_\gamma A_2(x_t)\label{dc9}~,
\end{eqnarray}
with $x_t=m_t^2/M_W^2$. 
The functions $A_1(x)$ and $A_2(x)$ are given by~\cite{chia}
\begin{equation}
A_1(x)=\frac{x}{2}\left[\frac{2x}{(1-x)^2} + 
\frac{(3-x)}{(1-x)^3}\ln x\right]~, 
\label{a1xt}
\end{equation} 
\noindent and
\begin{equation}
A_2(x)=-\frac{x}{4}\left[\frac{(1-5x)}{(1-x)^2} + 
\frac{(7-15x+4x^2)}{(1-x)^3}\ln x\right]~.
\label{a2xt}
\end{equation}
In the language of the effective hamiltonian formalism
these contributions translate into modifications of the matching conditions
for the Wilson coefficient functions $C_7$ and $C_9$ at the scale $M_W$. 
The first term in (\ref{bsgamma}) modifies 
$C_7(M_W)$ and therefore contributes to both $b\to s\gamma$ and 
$b\to s\ell^+\ell^-$, whereas the second term
only enters in the off-shell photon amplitude
and gives a contribution to $C_9(M_W)$. 
The anomalous TGC diagrams containing the $\Delta\kappa_\gamma$
have the same divergent structure as the SM TGC contributions, and therefore
the GIM mechanism renders them finite by decreasing their degree of 
divergence by one, thus 
eliminating an initially a logarithmic divergence.  

\begin{figure}
\center
\hspace*{0.7cm}
\psfig{figure=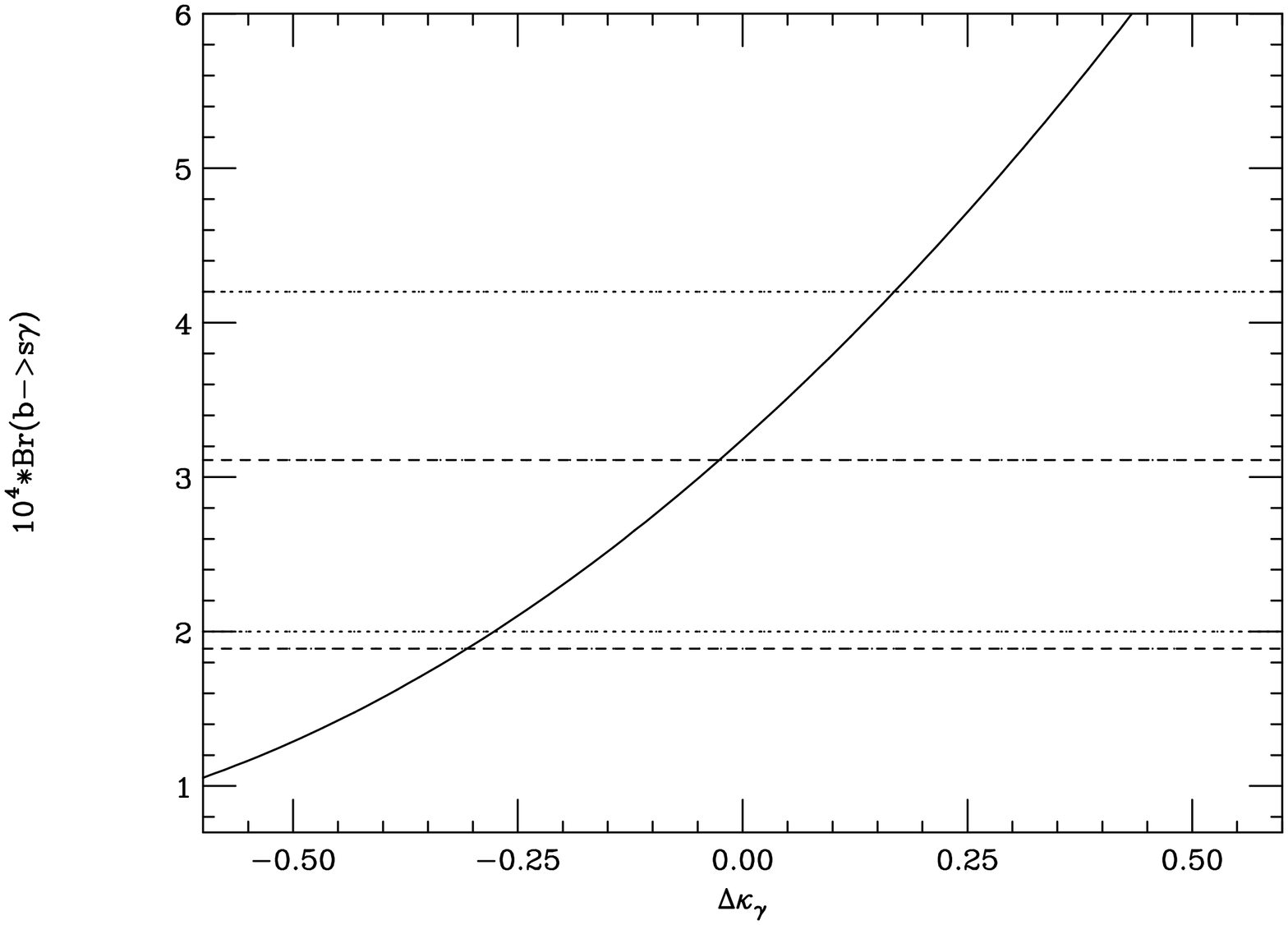,height=3.0in,angle=90}
\caption{\small\em The $Br(b\to s\gamma)$ vs. $\Delta\kappa_\gamma$. The 
dashed horizontal lines correspond to the $1\sigma$ CLEO 
measurement~{\rm\cite{cleo}}, whereas the dotted lines are the $1\sigma$ 
measurement from ALEPH~{\rm\cite{aleph}}.} 
\label{fig1}
\end{figure}

\noindent
In order to compute the effects in $B$ decays we evolve the Wilson 
coefficients
down to the scale $\mu\simeq m_b$ using standard procedures~\cite{heff}. 
In Fig.~\ref{fig1} we plot the $b\to s\gamma$ branching fraction  
as a function
of $\Delta\kappa_\gamma$. 
Also shown for reference are the $1\sigma$ intervals from the latest 
measurements of 
the CLEO collaboration~\cite{cleo}: 
$Br(b\to s\gamma)=(2.50\pm0.47\pm0.39)\times 10^{-4}$, 
as well as from the ALEPH collaboration~\cite{aleph}:
$Br(b\to s\gamma)=(3.11\pm0.80\pm0.72)\times 10^{-4}$. 
Combining these two results gives  
an approximate $1\sigma$ interval for $\Delta\kappa_\gamma$ 
$(-0.20,0.20)$.
Future measurements of the $b\to s\gamma$ 
branching ratio will greatly improve these constraints. 
For instance, 
a $20\%$ measurement of the $b\to s \gamma$ branching ratio 
would translate into the  
more stringent $1\sigma$ bound $-0.15<\Delta\kappa_\gamma <0.15$, 
if centered at the SM 
prediction.  

\begin{figure}
\center
\hspace*{0.7cm}
\psfig{figure=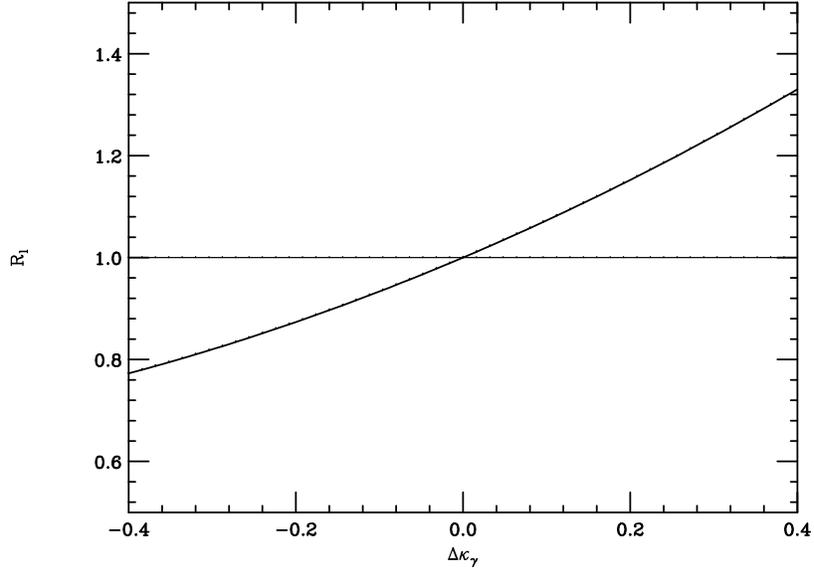,height=3.0in,angle=90}
\caption{\small\em The $b\to s\ell^+\ell^+$ branching 
ratio, normalized to its SM value, plotted  vs. $\Delta\kappa_\gamma$.} 
\label{fig2}
\end{figure}

\noindent
The dilepton modes $b\to s\ell^+\ell^-$ receive contributions 
from $\Delta\kappa_\gamma$ through both $\delta C_7$ and $\delta C_9$.  
In Fig.~\ref{fig2} the branching ratio $Br(b\to s \ell^+\ell^-)$, normalized 
by the SM value, is plotted versus $\Delta\kappa_\gamma$. Although the
sensitivity of these 
decay channels is similar to the one obtained in $b\to s\gamma$, 
the bounds are somewhat less stringent. 
This is more so when we consider that, unlike in $b\to s\gamma$, 
other anomalous TGC may significantly affect this amplitude. 
However, we will later come back to this point to show that it is 
possible to 
cleanly separate the contributions from the various relevant couplings 
even if 
only $b\to s\ell^+\ell^-$ decays are considered. 

\noindent
The sensitivity of these rare $B$ decays to $\Delta\kappa_\gamma$ is 
certainly 
comparable to that of higher energy experiments such as LEPII and the 
Tevatron. For instance the $95\%$~C.L. limits from LEPII~\cite{dkglep2} 
combining the data taken at $162$~GeV and at $172$~GeV ($10pb^{-1}$ at 
each energy)  are $(-1.10,1.80)$. On the other hand, the most 
recent measurements at the Fermilab Tevatron~\cite{dkgtev} put this coupling
in the range $(-0.36,0.45)$. The Tevatron bounds depend  on the 
scale of suppression introduced with the momentum dependence of the 
couplings, necessary to respect unitarity constraints. Both the Tevatron and 
the LEP bounds are obtained within a certain set of assumptions. 
Future LEPII measurements at higher energies, as well as Tevatron 
measurements with higher luminosity, will result in bounds  similar to 
the ones that will be obtained from FCNC processes named above.

We now turn to the effects of $\Delta g_1^Z$, an anomalous 
$W^+W^-Z$ coupling. Its presence affects the $b\to q \ell^+\ell^-$ amplitude
as well as the one of the neutrino modes $b\to q\nu\bar\nu$ and 
$K\to\pi\nu\bar\nu$. 
The modes governed by $b\to s\ell^+\ell^+$ are the most accessible
experimentally  
among the $B$ processes. 
Unlike the $\Delta\kappa_\gamma$ contribution, the diagrams
including $\Delta g_1^Z$ are still divergent, even after summing over 
all the 
intermediate up-quark states.  This divergence 
originates in the contributions from the longitudinal
pieces in the $W$ propagator 
and reflects the non-decoupling behavior of the Higgs sector. 
This logarithmic dependence of the loop effect on the high energy scale 
$\Lambda$  is a manifestation of the dynamics above this scale, and 
is presumably
related to electroweak symmetry breaking. Thus, this logarithmic
enhancement of the one-loop effect of $\Delta g_1^Z$ is of a rather 
fundamental 
origin~\cite{burgess} and makes FCNC particularly sensitive to this 
anomalous coupling.  

\begin{figure}
\center
\hspace*{0.7cm}
\psfig{figure=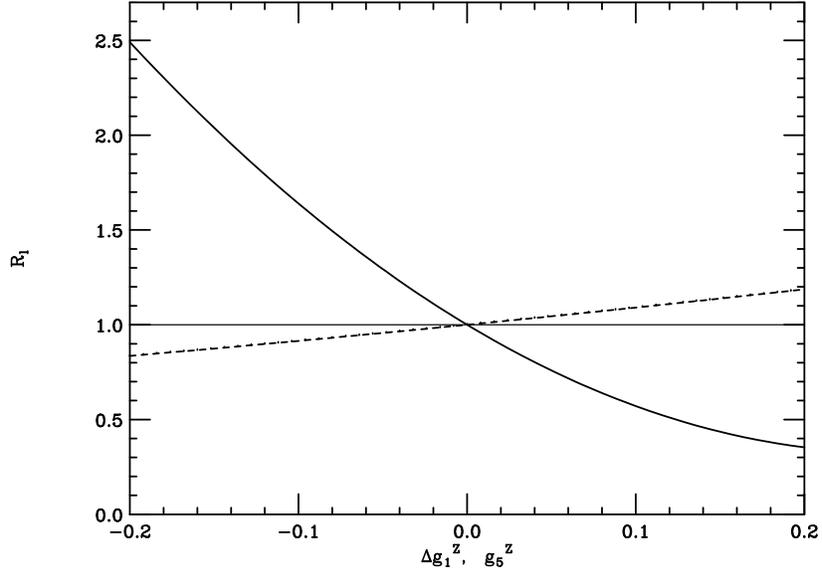,height=3.0in,angle=90}
\caption{\small\em 
The $b\to s\ell^+\ell^+$ branching 
ratio, normalized to its SM value, vs. $\Delta g_1^Z$ (solid line) and $g_5^Z$ 
(dashed line).} 
\label{fig3}
\end{figure}

\noindent 
The matching conditions for the Wilson coefficients
$C_9(M_W)$ and $C_{10}(M_W)$ are shifted by 
\begin{eqnarray} 
\delta C_9(M_W)&=&\Delta g_1^Z~\left(\frac{1-s^2\theta_w}
{s^2\theta_w}\right)~
(s^2\theta_w-\frac{1}{4})~B_1(x_t) \label{c9_b}\\
\delta C_{10}(M_W)&=&\Delta g_1^Z~\left(\frac{1-s^2\theta_w}
{s^2\theta_w}\right)~
\frac{B_1(x_t)}{4}\label{c10_b}~.
\end{eqnarray}
The function $B_1(x)$ is given simply by the leading logarithmic 
dependence, 
\begin{equation}
B_1(x)=\frac{3}{2}x\;\ln\frac{\Lambda^2}{M_W^2}+\dots~.
\label{b1x}
\end{equation}
In (\ref{b1x}), the dots stand for terms that are finite in the 
$\Lambda\to\infty$ limit. These terms are regularization scheme dependent
and, although formally subleading, could be numerically important. However, 
it is expected that the overall 
size of the effect is correctly estimated by the leading 
logarithmic behavior, 
barring precise cancellations with the finite terms. Thus, 
the results we present for $\Delta g_1^Z$ are meant to be 
indicative  of the sensitivity to this coupling but not a precise 
prediction\footnote{In reference~\cite{baillie}
this contribution 
was computed in the unitary gauge, and the scheme dependent terms were kept. 
Here we argue that only the logarithmic divergence can be trusted. The 
dependence on the scale $\Lambda$ is common to both results. }, 
something that cannot be achieved without knowledge of the full theory above
the energy scale $\Lambda$. 
The solid line in Fig.~\ref{fig3}, shows the branching ratio for 
$b\to s\ell^+\ell^-$, normalized to the SM model prediction,   
as a function of $\Delta g_1^Z$, 
where the high energy scale scale in (\ref{b1x}) is taken to be 
$\Lambda=2~$TeV.  
Although at present only upper limits on $b\to s\ell^+\ell^-$  processes
exist~\cite{uplim}, sensitivity to the SM predictions is expected to be 
achieved in the next round of $B$ physics experiments.  
For instance, measurements of 
$b\to s\ell^+\ell^-$ branching ratios with $30\%$ accuracy, 
can explore the region 
$|\Delta g_1^Z|<0.10$, a very competitive performance even when compared
with the high energy machines. For instance, LEPII is expected 
to just explore this region~\cite{lepfut}, 
whereas the Tevatron experiments, assuming an integrated luminosity of 
$1fb^{-1}$, will bound 
$\Delta g_1^Z$ to be in the interval $(-0.18,0.48)$~\cite{tevfut}. 
The main difference between these measurements and the FCNC decay modes
is that the latter have an additional dependence on $\Lambda$ from the 
logarithmic divergence. 

Next, we study 
the effects  of the $C$ and $P$ violating but 
$CP$ conserving coupling $g_5^Z$. 
These are simply obtained by the replacement 
$\Delta g_1^Z B_1(x_t) \to g_5^Z B_2(x_t)$ in eqns.~(\ref{c9_b}) 
and~(\ref{c10_b}), 
where  $B_2(x)$ is given by
\begin{equation}
B_2(x)=-\frac{3x}{1-x}\left(1+\frac{x\ln x}{1-x}\right)~.
\label{b2x}
\end{equation}
Unlike the contribution from $\Delta g_1^Z$, the resulting loop amplitude 
is finite, 
due to the fact that the $\epsilon_{\mu\nu\rho\sigma}$ tensor accompanying 
$g_5^Z$ does not couple to the longitudinal portion of the $W$ propagators.  
As a result, the contributions from this parameter to one-loop FCNC 
processes are not sensitive to the scale $\Lambda$. This, in turn, implies
that in this case there is no logarithmic enhancement as in the 
case of the $\Delta g_1^Z$ contribution and that these processes are not 
very sensitive to this coefficient, as can be seen from the 
dashed line in Fig.~\ref{fig3}. This is in agreement with the conclusions
of Ref.~\cite{sally}. 

\begin{figure}
\center
\hspace*{0.7cm}
\psfig{figure=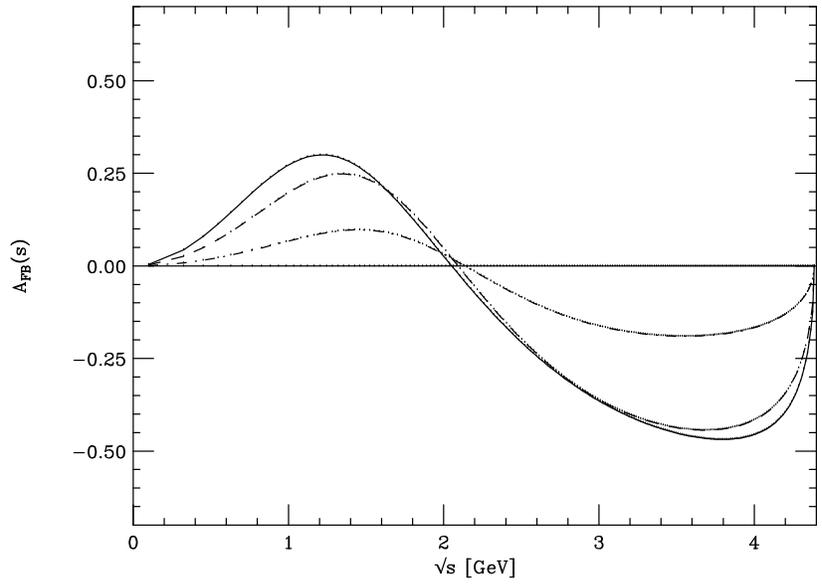,height=3.0in,angle=90}
\caption{\small\em 
The forward-backward asymmetry for leptons in $B\to K^*\ell^+\ell^-$, for 
$\Delta g_1^Z=0$, $0.1$ and $0.20$ (solid, dashed, dot-dashed respectively). 
Although these give large effects in the branching ratio,  
the position of the asymmetry zero is almost unaffected.}  
\label{fig4}
\end{figure}

From the above results, 
we conclude that $B$ decay processes involving one-loop FCNC 
are most sensitive to two CP conserving anomalous TGC, namely 
$\Delta\kappa_\gamma$ and $\Delta g_1^Z$. 
As we will see below, the analogous $K$ decay
modes have a similar sensitivity to $\Delta g_1^Z$. 
This is an important difference with the 
the high energy searches for these effects, where
the experiments are sensitive to several parameters giving room to possible
cancellations with the consequent weakening of the bounds. 
The limited sensitivity of the low energy FCNC transitions permits the clean
identification of the anomalous TGC. 
The obvious example is the fact that $b\to s\gamma$ is sensitive only
to $\Delta\kappa_\gamma$, among the CP conserving couplings. 
However, even when only considering $b\to s\ell^+\ell^-$, 
processes, where both $\Delta\kappa_\gamma$ and $\Delta g_1^Z$ contribute, 
it is possible to 
separate their effects.
This results from a very distinct pattern of shifts of the 
short distance Wilson coefficients. As it can be seen in (\ref{c9_b}), 
the shift in the coefficient $C_9(M_W)$ will be negligible due to 
the suppression factor $(\sin^2\theta_w-1/4)$, whereas this is not 
the case for $C_{10}(M_W)$. 
This is reflected in Fig.~\ref{fig4}, where we plot the 
forward-backward asymmetry
for leptons in $B\to K^*\ell^+\ell^-$ as a function of the dilepton 
mass. The asymmetry has a zero the position of which depends on the 
values of $C_7$ and $C_9$, but not on $C_{10}$~\cite{bkst}. 
Thus, values of $\Delta g_1^Z$ 
resulting in large deviations of the branching fractions in 
$b\to s\ell^+\ell^-$ decays, do not change the position of the 
asymmetry zero.
On the other hand, non-zero values of $\Delta\kappa_\gamma$ affect 
both $C_7$ and $C_9$
shifting the position where the asymmetry vanishes. In this way the angular 
information makes 
possible the separation  between $\Delta\kappa_\gamma$ and $\Delta g_1^Z$ 
effects that otherwise
could be unresolvable in the branching  ratio or even in the dilepton mass 
distribution.

\subsection{Rare $K$ Decays}

\begin{figure}
\center
\hspace*{0.7cm}
\psfig{figure=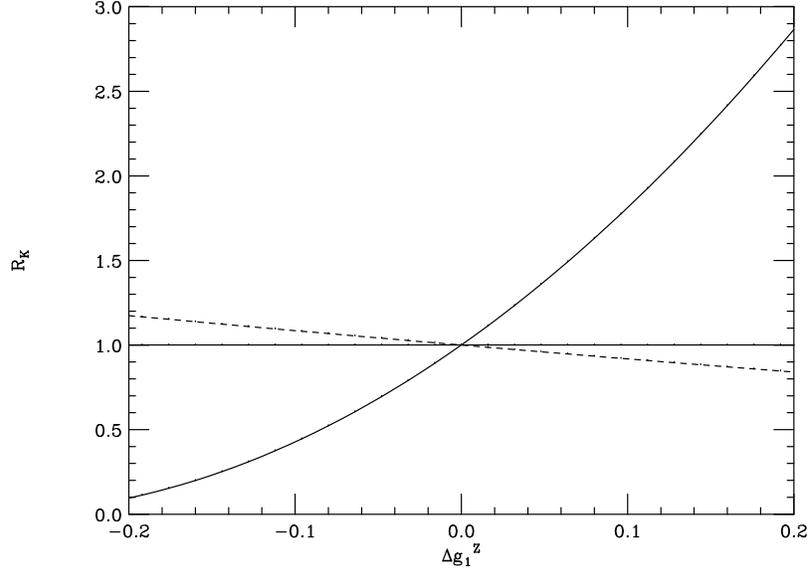,height=3.0in,angle=90}
\caption{\small\em The $K^+\to\pi^+\nu\bar\nu$ branching ratio, normalized to 
the SM prediction, plotted vs. the anomalous $WWZ$ couplings 
$\Delta g_1^Z$~(solid line), and $g_5^Z$ (dashed line). 
}
\label{fig5}
\end{figure}

Effects similar to those discussed above for $B$ decays are present in 
the analogous $K$ processes, due to the one loop contributions to
the $s\to d\gamma$ and $s\to d Z$ vertices. The photon mediated
transitions, such as $K\to\pi\ell^+\ell^-$ and hyperon radiative decays, 
are largely affected by long distance contributions which are 
theoretically uncertain and
make difficult the extraction of interesting short distance information.
On the other hand, $s\to d\nu\bar\nu$ transitions such as 
$K^+\to\pi^+\nu\bar\nu$ and $K_L\to\pi^0\nu\bar\nu$ are theoretically 
cleaner.
There are  two diagrams contributing to these processes, the box and 
the $s\to dZ$ penguin. The latter is sensitive to $\Delta g_1^Z$ and 
$g_5^Z$. 
The anomalous contribution to the $s\to d\nu\bar\nu$ amplitude 
can be written as 
\begin{equation}
\delta
{\cal A}(s\to d\nu\bar\nu) = \frac{4G_F}{\sqrt{2}}\frac{\alpha\cot^2\theta_w
}{8\pi}~V^*_{td}V_{ts}~\left(\Delta g_1^Z~B_1(x_t) + g_5^Z~B_2(x_t)\right)~
(\bar d_L\gamma_\mu s_L)(\bar{\nu}_L\gamma^\mu\nu_L)~,
\label{ktgc}
\end{equation}
with the functions $B_1(x)$ and $B_2(x)$ defined in (\ref{b1x}) and 
(\ref{b2x}). 
As discussed in the previous section, only the 
effect of $\Delta g_1^Z$ is sensitive to the logarithmic dependence on the 
high energy scale $\Lambda$, due to its coupling to the  
longitudinal gauge bosons. In Fig.~\ref{fig5} we plot the branching fraction 
for $K^+\to\pi^+\nu\bar\nu$, normalized to the SM expectation, as a function
of $\Delta g_1^Z$.  We observe that this decay mode has a sensitivity 
to $\Delta g_1^Z$ comparable to that of the $b\to s\ell^+\ell^-$ decays. 
However, the effect here is anti-correlated with 
the analogous one in $B$ processes.
Currently, this branching ratio is measured to be~\cite{kexp}
$Br(K^+\to\pi^+\nu\bar\nu)=(4.2^{+9.7}_{-3.5})\times 10^{-10}$, 
whereas the SM prediction is in the range $(0.60-1.00)\times 10^{-10}$
~\cite{kpism}. Thus, as it can be seen in Fig.~\ref{fig5}, there is room for 
relatively large values of $\Delta g_1^Z$ in both $B$ and $K$ FCNC decays.

\subsection{CP Violating Anomalous TGC}
In this section we discuss the possible effects of CP violating 
TGC. The most general form of the CP violating couplings of a neutral
gauge boson $N=\gamma,~Z$ to a $W$ pair is 
\begin{eqnarray}
{\cal L}_{\rm CPV} &=&g_{WWN}\left\{i\tilde{\kappa}_N~W^\dagger_\mu 
W_\nu~\tilde{N}^{\mu\nu} 
-g_4^N~W^\dagger_\mu W_\nu\left(\partial^\mu N^\nu
+\partial^\nu N^\mu\right)\right.\nonumber\\
& &\left. +i\frac{\tilde{\lambda}_N}{M_W^2}~W^\dagger_{\lambda\nu}
W^\mu_\nu\tilde{N}^{\nu\lambda}\right\}~,
\label{lcpv}
\end{eqnarray} 
with $\tilde{N}^{\mu\nu}=\frac{1}{2}\epsilon^{\mu\nu\alpha\beta}
N_{\alpha\beta}$. The effects of the $ZW^+W^-$ CP violating couplings 
in rare $B$ and $K$ decays are suppressed by powers of the typical external
momentum divided by $m_Z^2$, since all terms in (\ref{lcpv}) involve 
derivatives of the $Z$ field. On the other hand, the only
$\gamma W^+W^-$ coupling corresponding to a dimension four operator and
satisfying gauge invariance is 
$\tilde{\kappa}_\gamma$, since $\tilde\lambda_\gamma$ corresponds to a 
dimension six operator in the non-linear realization.
In the effective lagrangian (\ref{lnlo}) there are eight dimension four
operators contributing to the various CP violating terms in (\ref{lcpv}). 
The complete set of CP violating operators is 
given in the Appendix.
In that basis, the contributions to $\tilde\kappa_\gamma$ are
\begin{equation}
\tilde\kappa_\gamma= 2g^2\;\left(-\alpha_{16}
-4\alpha_{17} \right)~.
\label{kcpv}
\end{equation}
The $\tilde{\kappa}_\gamma$ contributions  to $b\to s \gamma$ 
and $b\to s\ell^+\ell^-$ take the 
form of complex shifts of the Wilson coefficients $C_7$ and $C_9$. 
The CP violating contribution to the coefficient of the magnetic moment 
operator $\bar s_L\sigma_{\mu\nu}b_R$ takes the form 
\begin{equation}
C_7(M_W)=C_7^{\rm SM}(M_W) -\frac{i}{2}\tilde\kappa_\gamma ~A_1(x_t)~,
\label{cpvc7}
\end{equation}
where the function $A_1(x)$ is given in equation (\ref{a1xt}). 
On the other hand, the leading order 
contributions from $\tilde\kappa_\gamma$ to the second term in equation 
(\ref{bsgamma}) corresponding to the shift in the coefficient $C_9(M_W)$, 
are of order ${\cal O}(m_b^2/M_W^2)$ and therefore negligible. 

\begin{figure}
\center
\hspace*{0.7cm}
\psfig{figure=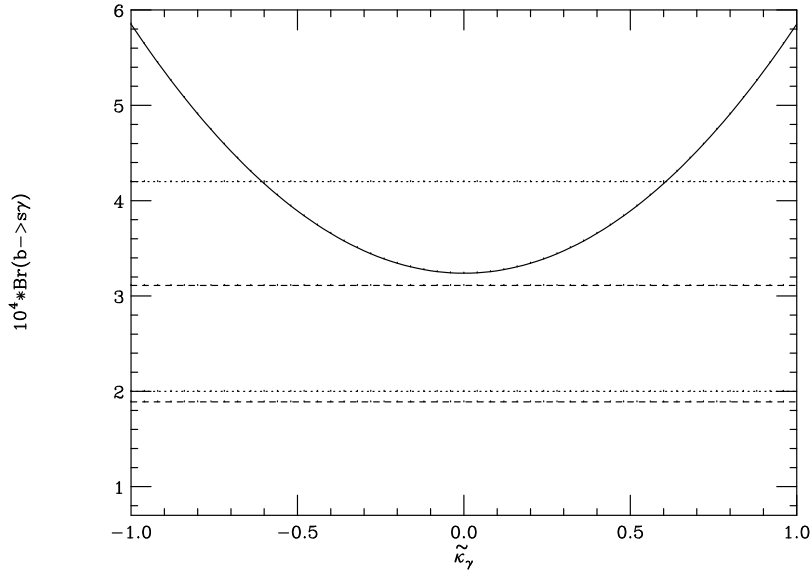,height=3.0in,angle=90}
\caption{\small\em 
The $Br(b\to s\gamma)$ vs.  the CP violating 
$WW\gamma$ coupling $\tilde\kappa_\gamma$. 
The 
dashed horizontal lines correspond to the $1\sigma$ CLEO 
measurement~{\rm\cite{cleo}}, whereas the dotted lines are the $1\sigma$ 
measurement from ALEPH~{\rm\cite{aleph}}.} 
\label{fig6}
\end{figure}

\noindent 
The $\tilde\kappa_\gamma$ contribution to $C_7(M_W)$ results always in a 
constructive effect in the $b\to s \gamma$ branching ratio, since there is no
interference with the SM. This translates into a rather tight bound on 
$\tilde\kappa_\gamma$, as it can be seen from Fig.~\ref{fig6}. 
Taking the $95\%$~C.L.
upper bound from the CLEO result, for instance, constrains this coupling
to be in the range
\begin{equation}
-0.60 ~\leq ~\tilde\kappa_\gamma ~\leq ~0.60 ~.
\label{range}
\end{equation}
The $b\to s \ell^+\ell^-$ modes give looser bounds. 
More stringent bounds than these come from the upper limits on the 
EDM of the neutron~\cite{edm}, giving the constraint 
$|\tilde\kappa_\gamma|< (2-3)\times 10^{-4}$. The neutron EDM bound is  
sensitive to the cutoff $\Lambda$ in the same way the $\Delta g_1^Z$ 
contributions
to rare $B$ and $K$ decays are. On the other hand, the present bounds are
cutoff independent by virtue of the GIM cancellation.   
Direct limits at hadron colliders are similar to the ones to be obtained
in $b\to s\gamma$. For instance, in Ref.~\cite{tevcpv} is is estimated that
the Tevatron with an integrated luminosity
of $1fb^{-1}$, will result in the bound $|\tilde\kappa_\gamma|<0.33$.

\noindent  
Taking into account the 
bound from (\ref{range}), we now consider possible CP violating observables. 
In the SM, CP violating asymmetries in $b\to d\gamma$ and 
$b\to d\ell^+\ell^-$
are expected to be in the few percent range~\cite{cpvsm}. On the other hand, 
they are negligibly small in the corresponding $b\to s$ transitions, due
to an extra factor of the Cabibbo angle. Thus, processes with 
strange mesons, 
such as $B\to K\ell^+\ell^-$, are free of SM sources of CP violation.  
For a partial rate asymmetry to arise, it is necessary that a CP-invariant
phase be present in the amplitude. In the case of $b\to s$ transitions this 
is 
provided, for instance, by the imaginary part of the one-loop insertion 
of the four-fermion operators such as  
$(\bar s_L\gamma_\mu c_L)(\bar c_L\gamma^\mu b_L)$
in the $b \to s\gamma^{(*)}$ vertex. The mixing of this operator with 
${\cal O}_9$ results in~\cite{heff}
\begin{equation}
C_9^{\rm eff.} = C_9(m_b) +  g(s)\;\left(3C_1+C_2+3C_3+C_4+3C_5+C_6
\right)~,
\label{c9eff}
\end{equation}
where the coefficients of the 
four-quark operators can be found in reference~\cite{heff} and the 
function $g(s)$ is given by
\begin{eqnarray}
g(s)&=&-\frac{4}{9}\ln z^2+\frac{8}{27}+\frac{16}{9}\frac{z^2}{s}\nonumber\\
& &-\frac{2}{9}\left(2+\frac{4z^2}{s^2}\right)
\left\{ 
\begin{array}{c}
2\,\sqrt{4z^2/s^2-1}\;\arctan(\frac{1}{\sqrt{z-1}})~~~~~~~~~~~~, 
{\rm for~}s<4m_c^2 \\
\\
\sqrt{1-4z^2/s^2}
\left[\ln\left(\frac{1+\sqrt{1-4z^2/s^2}}{1-\sqrt{1-4z^2/s^2}}
\right)+i\pi\right] ~~, {\rm for~}s>4m_c^2
\end{array}
\right. 
\label{gzs}
\end{eqnarray} 
where $z=m_c/m_b$. The imaginary part present in (\ref{gzs}), in 
combination with the CP violating phase coming from $\tilde\kappa_\gamma$, 
results in a small CP asymmetry.
For instance, when the constraint from equation~(\ref{range}) is considered, 
the partial rate
CP asymmetries in $B\to K\ell^+\ell^-$ are bound to be
\begin{equation}
A_{\rm CP}(B\to K\ell^+\ell^-)=\frac{\Gamma(B^+\to K^+\ell^+\ell^-) -
\Gamma(B^-\to K^-\ell^+\ell^-)}{\Gamma(B^+\to K^+\ell^+\ell^-) +
\Gamma(B^-\to K^-\ell^+\ell^-)} ~\simlt ~1\%~. 
\label{acpbs}
\end{equation}
Similar asymmetries are obtained in other strange meson modes, such as 
$K^0$, $K^*$,etc.  
Given the suppression of the SM asymmetries  observation of 
CP violation at this level would indicate the presence of new physics. 
On the other hand, 
several thousand reconstructed events would be needed for a significant 
measurement. 
We therefore conclude that the experimental observation of the effects 
of CP violating anomalous TGC is beyond the capabilities of  
first generation $B$ factories, where only a few hundred events 
are expected in these decay channels.

\section{Conclusions}

We have carried out a comprehensive study of the effects of anomalous
TGC in FCNC $B$ and $K$ decays. We have seen that these processes are 
sensitive to two $CP$ conserving couplings, $\Delta\kappa_\gamma$ 
and $\Delta g_1^Z$, as well as
to the CP violating coupling $\tilde\kappa_\gamma$. 
The reach of the next round of measurements at $B$ physics experiments 
such as Babar, Belle, CDF and D0 will put bounds on the CP conserving
couplings that are comparable to the limits to be obtained from direct 
gauge boson production at LEPII and an upgraded Tevatron. 
For comparison, in Table~I 
we quote the $95\%$ C.L. bounds on $\Delta\kappa_\gamma$ and $\Delta g_1^Z$ 
projected for 
LEPII~\cite{lepfut} at $190~$GeV and with $500~fb^{-1}$ of integrated 
luminosity, as well
as the limits for an upgraded Tevatron~\cite{tevfut} with $1~fb^{-1}$. 

\noindent 
For the future bounds from FCNC $B$ decays, we use very conservative estimates
of $1\sigma$ bounds that include {\em current} theoretical uncertainties 
present in the calculation of these modes. For instance, as mentioned earlier
and can be seen from Fig.~\ref{fig1}, 
a $20\%$ measurement of the $b\to s\gamma$ branching ratio would bound 
$\Delta\kappa_\gamma$
to be in the range $(-0.15,0.15)$.  For the bounds on $\Delta g_1^Z$, 
we rely on the projections for various $b\to s\ell^+\ell^-$ decay modes
to be observed at $B$ experiments at the SM level. 
For instance, 
several hundred events in the $B\to K^*\ell^+\ell^+$ channel 
will be available at the Tevatron experiments in the incoming run. 
This will allow not only 
tight bounds from the effect in the rate (Fig.~\ref{fig3}) but also 
the clean separation of the $\Delta g_1^Z$ coupling from possible effects 
from 
anomalous $WW\gamma$ couplings by analyzing dilepton angular information. 
The forward-backward asymmetry 
for leptons shown in Fig.~\ref{fig4} is an example of this separation: 
the position of the asymmetry zero is immune to $\Delta g_1^Z$, 
whereas it is very
sensitive to changes in the $WW\gamma$ couplings. 
On the other hand, it is possible to extract 
the short distance information from these exclusive modes  
by using a variety of techniques mostly related to heavy and light 
quark symmetry arguments, and with relatively small hadronic uncertainties
~\cite{bkst,afb}. 

\noindent
The limits on $\Delta g_1^Z$ can be further improved by future 
measurements of 
the $K^+\to\pi^+\nu\bar\nu$ branching fraction. This mode is as sensitive 
to the $WWZ$ anomalous coupling as the $b\to s\ell^+\ell^-$ modes, with the 
advantage that it is not polluted by the $WW\gamma$ couplings.

\phantom{xxxx}\vspace{0.1in}
\begin{center}
\begin{tabular}{|l||c|c|c|}
\hline
\multicolumn{4}{|c|}{Table~I.~{Comparison of bounds on Anomalous TGC. }}\\
\hline\hline
 & LEPII & Tevatron RunII & FCNC \\ 
 & 190~GeV & 1~fb$^{-1}$ & Decays\\ \hline
$\Delta\kappa_\gamma$ & (-0.25,0.40) & (-0.38,0.38) & (-0.20,0.20) \\ 
$\Delta g_1^Z$ & (-0.08,0.08) & (-0.18,0.48) & (-0.10,0.10) \\ 
$\tilde\kappa_\gamma$ & - & (-0.33,0.33) & (-0.50,0.50) \\ \hline
\end{tabular}
\end{center}
\phantom{xxxx}\vspace{0.1in}

\noindent 
We have also studied 
the effects of CP violating anomalous TGC, among which only 
$\tilde\kappa_\gamma$ is of relevance in FCNC decays. As seen in 
Fig.~\ref{fig6}, the current $1\sigma$ bound from the $b\to s\gamma$ branching 
ratio measurement is $-0.60<\tilde\kappa_\gamma<0.60$. Thus, the range quoted
in Table~I is a rather conservative estimate of what can be achieved by the 
next generation measurements of this decay channel.  
It compares well with what can be obtained by direct measurements, for 
instance, through $W\gamma$ production at the Tevatron~\cite{tevcpv}.  

\noindent
On the other hand, we have seen that the identification of an effect in 
the  radiative channels as coming from a CP violating coupling would require
measurements of CP asymmetries below $1\%$. This can only be obtained with 
several thousand reconstructed events in channels such as 
$B\to K\ell^+\ell^-$,
a goal that is beyond the first generation $B$ factories and perhaps to be 
attained by future dedicated $B$ experiments  at hadron colliders, such 
as the LHC-B at CERN or BTeV at the Tevatron.

We now briefly discuss the potential impact of these bounds on our 
understanding of the EWSB sector of the SM. As mentioned earlier, we focused 
on the non-linear realization of the EWSB sector, which is the appropriate
description in the absence of scalars with masses below the cutoff $\Lambda$. 
Within this framework the anomalous TGC $\lambda_\gamma$ and $\lambda_Z$
vanish at next-to-leading order in the effective theory (\ref{lnlo}), since 
they correspond to operators that are suppressed by  
$v^2/\Lambda^2$ relative to the dimension four set $\{{\cal L}_i\}$.  
The only consequence this power counting has in the analysis of 
low energy signals such as FCNC $B$ and $K$ decays, is the vanishing 
of the $\lambda_\gamma$ contributions, since the $\lambda_Z$ effects are 
suppressed by the factor $q^2/M_Z^2$ and are therefore negligible in any
description of the EWSB sector. Thus, as far as the anomalous $WWZ$ couplings
are concerned, the present analysis is valid in both the linear and 
non-linear  realizations. 

\noindent 
The effects of the coupling
$\Delta g_1^Z$ in FCNC processes are  enhanced by a 
logarithmic dependence on the high energy scale $\Lambda$. 
In the effective field theory language this leading logarithm  
coexists with 
finite counterterms which are naturally of comparable size.
As discussed in Section~3 in relation to
eqn.~(\ref{b1x}), the finite counterterms are model-dependent 
whereas the coefficient of the leading logarithm is determined 
at low energies independently of the specific theory above the scale 
$\Lambda$. Thus, although not the full answer, the logarithmic 
dependence provides us with the correct size of the effect, 
implying that the 
limits on $\Delta g_1^Z$ should be considered rough estimates 
of the effects, 
designed to evaluate the sensitivity of a given experiment to this 
physics.    
Furthermore, this logarithmically divergent behavior with the 
scale $\Lambda$   
arises as a consequence of the contributions of longitudinal 
components of $W^\pm$ in the loops, and is a manifestation of the 
non-standard behavior of the NGB of the electroweak symmetry breaking. 
All other anomalous TGC  give finite one-loop contributions to FCNC processes
due to the GIM mechanism and the fact that they only couple
to the transverse piece of the gauge boson propagators. The GIM cancellation 
ensures that the bounds obtained on $\Delta\kappa_\gamma$ and $g^Z_5$
are more precise. 
Therefore the bounds from rare $B$ and $K$ decays  
on $\Delta g_1^Z$ will be less precise
(perhaps good up to factors of two or so), but is the coupling to which FCNC 
$B$ and $K$ 
decays are most sensitive and potentially the most interesting one.

\noindent
With respect to the expected size of the effects, we emphasize that the 
present study is model-independent and that to compute the 
coefficients $\{\alpha_i\}$ of the effective lagrangian (\ref{lnlo})
knowledge of the full theory above the matching scale $\Lambda$ is needed. 
However, it is possible to apply dimensional arguments to these couplings. 
For instance, naive dimensional analysis (NDA) \cite{nda} suggests that
\begin{equation}
\alpha_i\simeq {\cal O}(1)\times\frac{v^2}{\Lambda^2}~,
\label{da}
\end{equation}
with the scale of new physics obeying $\Lambda\simlt 4\pi v$.
However, in practice this power counting can only be applied to those 
coefficients
that respect the custodial $SU(2)$ symmetry that ensures that 
$\Delta\rho_*=\alpha T$
is small compared to one. 
As discussed in Ref.~\cite{sally}, this constraint implies that 
custodial breaking terms in ${\cal L}_{\rm eff.}$ should naturally be 
further suppressed by an extra factor of ${\cal O}(10^{-2})$ or so. 
In terms of the anomalous TGC this means that it is natural to expect that 
$g_5^Z$ is no larger than ${\cal O}(10^{-4}-10^{-3})$
On the other hand, $\Delta\kappa_\gamma$ and $\Delta g_1^Z$ 
receive contributions
from custodial conserving terms and then 
are expected to be in the ${\cal O}(10^{-3}-10^{-1})$ range in these 
scenarios.
A sizeable fraction of this range can be reached by FCNC processes, 
which are 
sensitive to anomalous TGC as small as a few percent. 
For the coefficient $\Delta g_1^Z$ this is true 
even in  the first generation of $B$ factory experiments and for  
$\simeq 30\%$ measurements 
of the $K^+\to\pi^+\nu\bar\nu$ branching ratio. 
For both the CP conserving $\Delta\kappa_\gamma$ and CP violating 
$\tilde\kappa_\gamma$ 
$WW\gamma$ anomalous couplings, a few percent precision will only be achieved
with at least one order of magnitude more reconstructed events, 
to be available at the proposed LHC-B and BTeV experiments.

\vskip1.0cm
\noindent
{\bf Acknowledgments}

\noindent
The author thanks Kaoru Hagiwara for suggesting to look at the 
CP violating TGC, and to Guo-Hong Wu for clarifying comments on the 
operator basis of Reference~\cite{appelquist}. 
Useful conversations with Dieter Zeppenfeld
are also acknowledged. 
This work was supported  by the U.S.~Department of Energy 
under  
Grant No.~DE-FG02-95ER40896 and the University of 
Wisconsin Research Committee with funds granted by the Wisconsin 
Alumni Research Foundation.

\newpage
\appendix
\section*{Appendix}
\setcounter{equation}{0}
\renewcommand{\theequation}{A.\arabic{equation}}
Here we specify the operator basis used for 
the effective lagrangian of the EWSB sector of the SM.  
Defining 
\begin{equation}
\begin{array}{lccr}
T\equiv U\tau^3 U^\dagger,~ &~~~~~~~&
~~~~~~~~~~& V\equiv(D_\mu U)U^\dagger,~
\end{array}
\label{deftv}
\end{equation}
with $U$ and the covariant derivative defined in Section~1, 
all operators up to dimension four that are invariant 
under $SU(2)_L\times U(1)_Y$ can be written in terms of the gauge
fields, $T$, $V_\mu$ and 
\begin{equation}
{\cal D}_\mu {\cal O}\equiv \partial{\cal O}+ig[W_\mu,{\cal O}]~. 
\label{covder}
\end{equation}
The dimension two operator 
${\cal L}_1'=(v^2/4)\left[{\rm Tr}(TV_\mu)\right]^2$, gives a contribution
to the $T$ parameter and thus its coefficient is greatly 
constrained~\cite{appelquist}. 
The CP-invariant dimension-four operators of eqn.~(\ref{lnlo}) are given by 
\begin{eqnarray}
{\cal L}_1&=&\frac{1}{2}gg'B_{\mu\nu}{\rm Tr}(TW^{\mu\nu}) 
\nonumber\\
{\cal L}_2&=&\frac{1}{2}ig'B_{\mu\nu}{\rm Tr}(T[V^\mu,V^\nu])
\nonumber\\
{\cal L}_3&=&ig{\rm Tr}(W_{\mu\nu}[V^\mu,V^\nu]) 
\nonumber\\
{\cal L}_4&=&\left[{\rm Tr}(V_\mu V_\nu)\right]^2 
\nonumber\\
{\cal L}_5&=&\left[{\rm Tr}(V_\mu V^\mu)\right]^2 
\nonumber\\
{\cal L}_6&=&{\rm Tr}(V_\mu V_\nu){\rm Tr}(TV^\mu)
{\rm Tr}(TV^\nu)
\nonumber\\
{\cal L}_7&=&{\rm Tr}(V_\mu V^\mu){\rm Tr}(TV_\nu)
{\rm Tr}(TV^\nu) 
\label{opscpc}\\
{\cal L}_8&=&\frac{1}{4}g^2\left[{\rm Tr}(TW_{\mu\nu})\right]^2 
\nonumber\\
{\cal L}_9&=&\frac{1}{2}ig{\rm Tr}(TW_{\mu\nu}){\rm Tr}(T[V^\mu,V^\nu]) 
\nonumber\\
{\cal L}_{10}&=&\frac{1}{2}\left[{\rm Tr}(TV_\mu)
{\rm Tr}(TV_\nu)\right]^2
\nonumber\\
{\cal L}_{11}&=&{\rm Tr}\left[({\cal D}_\mu V^\mu)^2\right]
\nonumber\\
{\cal L}_{12}&=&{\rm Tr}(T{\cal D}_\mu {\cal D}_\nu V^\nu)
{\rm Tr}(TV^\mu)
\nonumber\\
{\cal L}_{13}&=&\frac{1}{2}\left[{\rm Tr}(T{\cal D}_\mu V_\nu)\right]^2
\nonumber\\
{\cal L}_{14}&=&g\epsilon^{\mu\nu\rho\sigma}{\rm Tr}(TV_\mu)
{\rm Tr}(V_\nu W_{\rho\sigma})~.
\nonumber
\end{eqnarray}
This CP-conserving basis  contains three additional operators
with respect to Reference~\cite{appelquist}. The operators 
${\cal L}_{11}$, ${\cal L}_{12}$ and ${\cal L}_{13}$ either vanish or can 
be written as linear combinations of the others in the limit of massless
fermions, in which  
${\cal D}_\mu V^\mu\simeq 0$. They are generally neglected when considering
on-shell amplitudes. However, here we will insert these operators in 
one loop processes.
Finally, there are three independent CP violating operators, 
as found in Reference~\cite{appelquist}. They are
\begin{eqnarray}
{\cal L}_{15}&=&g{\rm Tr}(TV_\mu){\rm Tr}(V_\nu W^{\mu\nu})
\nonumber\\
{\cal L}_{16}&=&gg'\epsilon^{\mu\nu\rho\sigma}B_{\mu\nu}
{\rm Tr}(TW^{\rho\sigma})
\label{opscpv}\\
{\cal L}_{17}&=&g^2\epsilon^{\mu\nu\rho\sigma}
{\rm Tr}(TW_{\mu\nu}){\rm Tr}(TW^{\rho\sigma}) 
\nonumber\\
{\cal L}_{18}&=&{\rm Tr}(V_\mu {\cal D}_\nu V^\nu){\rm Tr}(TV^\mu)
\nonumber\\
{\cal L}_{19}&=&{\rm Tr}\left(\left[V_\mu,T\right]
{\cal D}^\mu{\cal D}^\nu V_\nu\right)~.
\nonumber
\end{eqnarray}
Only ${\cal L}_{16}$ and ${\cal L}_{17}$ contain  $\tilde{F}_{\mu\nu}$
terms which then will contribute to $\tilde{\kappa}_\gamma$, as it
can be seen in eqn.~(\ref{kcpv}). The last two operators vanish in the 
limit  of massless fermions, in which case the CP violating basis coincides
with the one in Reference~\cite{appelquist}.


\newpage

\end{document}